\documentclass[11pt]{article}
\usepackage{float}
\floatstyle{boxed}
\pdfoutput = 1
\usepackage{epsfig}
\usepackage{graphics}
\usepackage{amsmath}
\usepackage{graphicx} 
\begin{document}
\title
{Black hole thermodynamics and generalized uncertainty principle with higher order terms in momentum uncertainty }
\author{
{\bf {\normalsize Sunandan Gangopadhyay}$^{a,d}$
\thanks{sunandan.gangopadhyay@gmail.com, sunandan.gangopadhyay@bose.res.in, sunandan@associates.iucaa.in}},\,
{\bf {\normalsize Abhijit Dutta}$^{b,c}
$\thanks{dutta.abhijit87@gmail.com}}\\
$^{a}$ {\normalsize Department of Theoretical Sciences, S. N. Bose National Centre for Basic Sciences,}\\
{\normalsize JD Block, Sector III, Saltlake, Kolkata 700106, West Bengal, India }\\[0.2cm]
$^{b}$ {\normalsize Department of Physics, West Bengal State University, Barasat, Kolkata 700126, India }\\[0.2cm]
$^{c}$ {\normalsize Department of Physics, Kandi Raj College, Kandi, Murshidabad 742137, India }\\[0.2cm]
$^{d}${\normalsize Visiting Associate in Inter University Centre for Astronomy \& Astrophysics (IUCAA),}\\
{\normalsize Pune 411007, India}\\[0.1cm]
}
\date{}

\maketitle
\begin{abstract}
	In this paper we study the modification of thermodynamic properties of Schwarzschild and Reissner-Nordstr\"{o}m black hole in the framework of generalized uncertainty principle with correction terms upto fourth order in momentum uncertainty. The mass-temperature relation and the heat capacity for these black holes have been investigated. These have been used to obtain the critical and remnant masses. The entropy expression using this generalized uncertainty principle reveals the area law upto leading order logarithmic corrections and subleading corrections of the form $\frac{1}{A^n}$.  The mass output and radiation rate using Stefan-Boltzmann law have been computed which show deviations from the standard case and the case with the simplest form for the generalized uncertainty principle.   
\end{abstract}
\section{Introduction}
\noindent The consistent unification of quantum mechanics (QM) with general relativity (GR) is one of the major task in theoretical physics. GR deals with the definition of world-lines of particles, which is in  contradiction with QM since it does not allow the notion of trajectory due to the presence of an uncertainty in the determination of the momentum and position of a quantum particle. It has been the aim to unify these two theories into one theory known as quantum gravity. It is quite interesting that all approaches towards quantum gravity such as black hole physics \cite{maggiore}-\cite{gross2}, string theory \cite{str,konishi} or even \textit{Gedanken} experiment \cite{scardigli} predict the existence of a minimum measurable length. The occurrence of such a minimal length also arises in various theories of quantum gravity phenomenology, namely, the generalized uncertainty principle (GUP) \cite{koni}-\cite{magg3}, modified dispersion relation (MDR) \cite{amelino2}-\cite{sadatian}, deformed special relativity (DSR) \cite{magueijo}, to name a few. It is now widely accepted that Heisenberg uncertainty principle would involve corrections from gravity at energies close to the Planck scale. Thus emergence of a minimal length seems to be inevitable when gravitational effects are taken into account. There has been a lot of work incorporating the existence of a minimal length scale in condensed matter and atomic physics experiments such as Lamb Shift, Landau levels and the scanning Tunneling Microscope \cite{das1}-\cite{das6}, loop quantum gravity \cite{gambini}-\cite{amelino5}, noncommutative geometry \cite{girelli}, computing Planck scale corrections to the phenomena of superconductivity and quantum Hall effect \cite{mann} and understanding its consequences in cosmology \cite{Das7,Das8}. 

The incorporation of the GUP to study black hole thermodynamics has been another interesting area of active research \cite{adler}-\cite{feng}. It has been observed that the GUP reveals a self-complete characteristic of gravity which basically amounts to hiding any curvature singularity behind an event horizon as a consequence of matter compression at the Planck scale \cite{dvali1}-\cite{isi}. Further, the effects of the GUP have also been considered  in the tunneling formalism for Hawking radiation to evaluate the quantum-corrected Hawking temperature and entropy of a Schwarzschild black hole \cite{nozari4}-\cite{yang}. In our earlier findings \cite{abhijit1}-\cite{abhijit4}, we have studied the modification of thermodynamic properties, namely the temperature, heat capacity and entropy of black holes due to the simplest form of the GUP. Interestingly the correction to the Schwarzschild black hole temperature
due to quadratic and linear-quadratic GUP has also been compared with the corrections from
the quantum Raychaudhuri equation \cite{alasfer}. Very recently the Lorentz-invariance-violating class of dispersion relations have been applied to study the thermodynamics of black holes \cite{li}. It would therefore be interesting to compare these results with those coming from the GUP.

The above studies motivate us to investigate the modification of thermodynamic properties for Schwarzschild and Reissner-Nordstr\"{o}m (RN) black holes using the form of the GUP proposed in \cite{rb}. This GUP involves higher order terms in the momentum uncertainty. We compute the remnant and critical masses analytically for these black holes below which the temperature becomes ill-defined. We then use the Stefan-Boltzmann law to estimate the mass and the energy output as a function of time. We finally compute the entropy and obtain the well known area theorem containing corrections from the GUP with higher order terms in momentum uncertainty. 

The paper is organized as follows. In section 2, we study the thermodynamics of Schwarzschild black hole taking into account the effect of the GUP, with higher order terms in momentum uncertainty. In sub-section 2.1, we also obtain the mass and radiation rate characteristics for the Schwarzschild black hole as a function of time by using the Stefan-Boltzmann law. In section 3, we study the thermodynamics of Reissner-Nordstr\"{o}m black holes taking into account the effect of the GUP. Finally, we conclude in section 4.

\section{Thermodynamics of Schwarzschild black hole}
In this paper we work with the following form of the GUP \cite{nozari6}
\begin{eqnarray}
\Delta x\Delta p&\geq&\frac{\hbar}{2}\sum_{i=0}^{\infty}a_i\left(\frac{l_p \Delta p}{\hbar}\right)^{2i};~~~~~~~~~~~~~~[ a_0 = 1; a_i>0, i=1,2,...]
\label{GUP in general}
\end{eqnarray}
where $l_p$ is the Planck length ($\sim 10^{-35}m$).
\noindent Keeping terms upto fourth order in momentum uncertainty, we have 
\begin{eqnarray}
\Delta x\Delta p&\geq&\frac{\hbar}{2}\left\{a_0 + a_1\left(\frac{l_p \Delta p}{\hbar}\right)^2 + a_2\left(\frac{l_p \Delta p}{\hbar}\right)^4\right\}.
\label{GUP}
\end{eqnarray}
We now consider a Schwarzschild black hole of mass $M$. In the vicinity of the event horizon of the black hole, let a pair (particle-antiparticle) production occur. For simplicity we consider the particle to be massless. The particle with negative energy falls inside the horizon and that with positive energy escapes outside the horizon and gets observed by some observer at infinity. The momentum of the emitted particle (p), which also characterizes the temperature (T), is of the order of its uncertainty in momentum  $\Delta p$. Consequently
\begin{eqnarray}
T=\frac{(\Delta p) c}{k_B}
\label{momentum uncertainty-temp. reln.}
\end{eqnarray}
where $c$ is the speed of light and $k_B$ is the Boltzmann constant.

\noindent The Hawking temperature of the black hole will be equal to the temperature of the particle when thermodynamic equilibrium is reached. 
The uncertainty in the position of a particle near the event horizon of the Schwarzschild black hole will be of the order of the Schwarzschild radius of the black hole
\begin{eqnarray}
\Delta x=\epsilon r_{s}~;~
r_{s}=\frac{2GM}{c^2}
\label{position uncertainty-mass reln.}
\end{eqnarray}
where $\epsilon$ is a calibration factor, $r_s$ is the Schwarzschild radius and $G$ is the Newton's universal gravitational constant.

\noindent To relate the Hawking temperature of the black hole with the mass of the black hole, we consider the saturated form of the GUP (\ref{GUP})
\begin{eqnarray}
\Delta x\Delta p = \frac{\hbar}{2}\left\{a_0 + a_1\left(\frac{l_p \Delta p}{\hbar}\right)^2 + a_2\left(\frac{l_p \Delta p}{\hbar}\right)^4\right\}
\label{saturaed form of GUP}.
\end{eqnarray}
Substituting eq.(s) (\ref{momentum uncertainty-temp. reln.}) and (\ref{position uncertainty-mass reln.}) in eq. (\ref{saturaed form of GUP}) gives
\begin{eqnarray}
M&=&\frac{{M_p}^2 c^2}{4\epsilon k_B T}\left\{1 + a_1\frac{k_B^2 T^2}{M_p^2c^4} +  a_2\frac{k_B^4 T^4}{M_p^4c^8}\right\}
\label{Mass-Temp Relation}
\end{eqnarray}
where the relations  $\frac{c\hbar}{l_p}=M_{p}c^{2}$ and $M_{p}=\frac{c^2 l_p}{G}$,  ($M_{p}$ being the Planck mass) have been used. 
\noindent In the absence of corrections due to GUP, eq.(\ref{Mass-Temp Relation}) reduces to
\begin{eqnarray}
M=\frac{M_{p}^2 c^2}{4\epsilon k_{B}T}~.
\label{GUP in absence of quantum gravity}
\end{eqnarray}
 The value of $\epsilon$ now gets fixed to $2\pi$ by comparing this expression with the semi-classical Hawking temperature  $T=\frac{M_{p}^{2}c^2}{8 \pi M k_{B}}$   \cite{hawk1,hawk2}.
\noindent The mass-temperature relation (\ref{Mass-Temp Relation}) finally takes the form 
\begin{eqnarray}
M&=& \frac{{M_p}^2 c^2}{8\pi}\left\{\frac{1}{k_B T} + a_1\frac{k_B T}{M_p^2c^4} +  a_2\frac{k_B^3 T^3}{M_p^4c^8}\right\}
\label{Mass-Temp Relation1}.
\end{eqnarray}
\noindent The heat capacity of the black hole therefore reads
\begin{eqnarray}
C&=&c^2\frac{dM}{dT}~
\label{heat capacity defn.}\\
&=&\frac{{M_p}^2 c^4}{8\pi}\left\{-\frac{1}{k_B T^2} + a_1\frac{k_B }{M_p^2c^4} +  3a_2\frac{k_B^3 T^2}{M_p^4c^8}\right\}\nonumber\\
&=& \frac{k_B}{8\pi}\left\{-\left(\frac{M_pc^2}{k_B T}\right)^2 + a_1 +  3 a_2\left(\frac{k_B T}{M_pc^2}\right)^2\right\}.
\label{heat capacity}
\end{eqnarray}

\noindent The above expression takes the form
\begin{eqnarray}
C = \frac{k_B}{8\pi}\left\{-\frac{1}{{T'}^2}+a_1+3a_2 {T'}^2\right\}
\label{heat capacity 2}
\end{eqnarray}
after the following notations are introduced
\begin{eqnarray}
M'=\frac{8\pi M}{M_p}~;~T'=\frac{k_B T}{M_p c^2}~.
\label{notation for mass and temp}
\end{eqnarray}

\noindent The mass of the black hole decreases due to radiation from the black hole. This leads to an increase in the temperature of the black hole. It can be observed from eq.(s) (\ref{GUP in absence of quantum gravity}) and (\ref{heat capacity}) that there exists a finite temperature at which the heat capacity vanishes. To find out this temperature, we set C = 0.  This gives
\begin{eqnarray}
3a_2 {T'}^4 + a_1 {T'}^2 -1 = 0.
\end{eqnarray}  
\noindent Solving this, we get
\begin{eqnarray}
T'^2 = \frac{1}{6a_2}\left\{-a_1 + \sqrt{{a_1}^2 + 12a_2}\right\}
\end{eqnarray}
\noindent where the positive sign before the square root has been taken so that the above result reduces to corresponding result when $a_2 = 0$ \cite{abhijit2}.
 
\noindent Finally we get the expression for $T'$  to be
\begin{eqnarray}
T' = \frac{1}{\sqrt{6a_2}}\left\{-a_1 + \sqrt{{a_1}^2 + 12a_2}\right\}^\frac{1}{2}
\label{maxm. temp.}.
\end{eqnarray}
\noindent 
\noindent Now in terms of $T', M'$ the mass-temperature relation (\ref{Mass-Temp Relation1}) can be represented as
\begin{eqnarray}
a_2 {T'}^4 + a_1 {T'}^2 - M'T' + 1 = 0
\label{Mass-Temp Relation2} .
\end{eqnarray}
\noindent The remnant mass can now be obtained by substituting eq.(\ref{maxm. temp.}) in eq.(\ref{Mass-Temp Relation2}). This yields
\begin{eqnarray}
M'_{rem} &=& \frac{1}{T'}\left\{a_2 {T'}^4 + a_1 {T'}^2 + 1\right\}\nonumber\\
\Rightarrow M_{rem} &=&\frac{M_p}{8 \pi}\frac{\sqrt{6a_2}}{9a_2\left\{-a_1 + \sqrt{{a_1}^2 + 12a_2}\right\}^\frac{1}{2}}\left[-{a_1}^2 + a_1 \sqrt{{a_1}^2 + 12a_2} + 12a_2\right].
\label{Remnant Mass}
\end{eqnarray}
\noindent Reassuringly the above result reduces to the result in   $a_2\rightarrow 0$ limit \cite{abhijit2}
\begin{eqnarray}
 M_{rem} = \frac{M_p \sqrt{a_1}}{4 \pi}
\label{Remnant Mass for $a_2 = 0$}~.
\end{eqnarray}
\noindent Now for $ a_1\rightarrow0$ , $a_2\neq 0$,  the remnant mass is given by
\begin{eqnarray}
M_{rem} = \frac{M_p}{6 \pi}(3a_2)^\frac{1}{4}
\label{Remnant Mass for $a_1 = 0$}~.
\end{eqnarray}
\noindent Also for $ a_1\rightarrow0$, the mass-temperature relation (\ref{Mass-Temp Relation2}) reads 
\begin{eqnarray}
{T'}^4-\frac{M'}{a_2}T'+\frac{1}{a_2} = 0
\label{Mass-Temp Reln. for $a_1 = 0$}~.
\end{eqnarray} 

\noindent The solution of this bi-quadratic equation in $T'$ yields 
\begin{eqnarray}
T'=\left[\frac{{M'}^2}{16{a_2}^2}+\sqrt{\frac{{M'}^4}{256{a_2}^4}-\frac{1}{27{a_2}^3}}\right]^{\frac{1}{3}}+\left[\frac{{M'}^2}{16{a_2}^2}-\sqrt{\frac{{M'}^4}{256{a_2}^4}-\frac{1}{27{a_2}^3}}\right]^{\frac{1}{3}}~.
\end{eqnarray}
The above relation readily implies the existence of a critical mass
below which the temperature will be a complex quantity
\begin{eqnarray}
M_{cr} = \frac{M_p}{6 \pi}(3a_2)^\frac{1}{4}
\label{Critical Mass for $a_1 = 0$}~.
\end{eqnarray}
This demonstrates that the remnant and critical masses are equal.

\noindent At this point, we would like to make a comment. It can be observed from the above analysis that analytical expressions for the remnant and critical masses can be obtained even if one retains terms of order of $(\Delta p)^8$ in the momentum uncertainty. This is because it leads to an equation of the form $a~{T'}^8 + b~{T'}^6 + c~{T'}^4 + d~{T'}^2 + e=0$ when the condition $C=0$ is imposed. This equation can be solved analytically to obtain the remnant mass. If we keep terms beyond this order in momentum uncertainty, analytical expressions for the remnant and  critical masses can not be obtained.

\noindent The black hole entropy from the first law of black hole thermodynamics is given by
\begin{eqnarray}
S &=& \int c^2 \frac{dM}{T} = \int C\frac{dT}{T}\nonumber\\
\label{entropy defn}
&=& \int \frac{k_B}{8\pi}\left[-\left(\frac{M_p c^2}{k_B T}\right)^2 + a_1 + 3a_2\left(\frac{k_B T}{M_p c^2}\right)^2\right]\frac{dT}{T}\nonumber\\
&=& \frac{k_B}{8\pi}\left[\frac{1}{2}\left(\frac{M_p c^2}{k_B T}\right)^2 + a_1 \ln {\left(\frac{k_B T}{M_p c^2}\right)} +  \frac{3a_2}{2}\left(\frac{M_p c^2}{k_B T}\right)\right] \nonumber\\
&=& \frac{k_B}{8\pi}\left[  \frac{1}{2{T'}^2} + a_1 \log T' + \frac{3}{2}a_2 {T'^2}\right].
\label{entropy}
\end{eqnarray}

\noindent To obtain the entropy S in terms of the mass M of the black hole, we need to consider eq.(\ref{Mass-Temp Relation2}) to obtain an expression for the temperature T in terms of mass M. 

\noindent Eq.(\ref{Mass-Temp Relation2}) yields upto $\mathcal{O}( {a_1}^2, {a_2}^2 ,a_1 a_2)$

\begin{eqnarray}
 T' = \frac{1}{M'} + \frac{a_1}{{M'}^3} + \frac{a_2}{{M'}^5} + \frac{2{a_1}^2}{{M'}^5} + \frac{6 a_1 a_2}{{M'}^7} + \frac{4{a_2}^2}{{M'}^9}.
\label{Mass-Temp Relation3}
\end{eqnarray}
\noindent Now the entropy expression in terms of the  mass can be written as 
\begin{eqnarray}
\frac{S}{k_B} &=& \frac{1}{8\pi}\Bigg[\frac{1}{2}{{M'}^2}-a_1-a_1\log M'+\frac{{a_1}^2}{2{M'}^2}+\frac{a_2}{2{M'}^2}+\frac{a_1 a_2}{{M'}^4}+\frac{{a_2}^2}{2{M'}^6}\Bigg]+\mathcal{O}({a_1}^2 a_2 , a_1 {a_2}^2 , {a_1}^3,{{a_2}^3 })\nonumber\\
 &=&  \frac{S_{BH}}{k_B}-\frac{a_1}{16\pi}\log\left(\frac{S_{BH}}{k_B}\right)-\frac{a_1}{16\pi}\log(16\pi)-\frac{a_1}{8\pi}+\frac{{a_1}^2 +a_2}{(16\pi)^2}\left(\frac{S_{BH}}{k_B}\right)^{-1}\nonumber\\&&
+\frac{2a_1  a_2}{(16\pi)^3}\left(\frac{S_{BH}}{k_B}\right)^{-2}+ \frac{{a_2}^2 }{(16\pi)^4}\left(\frac{S_{BH}}{k_B}\right)^{-3}+\mathcal{O}({a_1}^2 a_2 , a_1 {a_2}^2 , {a_1}^3,{{a_2}^3 })
\end{eqnarray}
where   $\frac{S_{BH}}{k_B} =\frac{4 \pi M^2}{M_p^2}$ is the semi-classical 
Bekenstein-Hawking entropy for the Schwarzschild black hole.
\noindent  In terms of the black hole horizon area $A=4\pi r_{s}^2 =16\pi \frac{G^2 M^2}{c^4} = 4 l_p^2 \frac{S_{BH}}{k_B}$,
the above entropy expression can be written as
\begin{eqnarray}
\frac{S}{k_B} &=&  \frac{A}{4l_p^2}-\frac{a_1}{16\pi}\log\left(\frac{A}{4l_p^2}\right)-\frac{a_1}{16\pi}\log(16\pi)-\frac{a_1}{8\pi}+\frac{{a_1}^2 +a_2}{(16\pi)^2}\left(\frac{A}{4l_p^2}\right)^{-1}\nonumber\\&&
+\frac{2a_1  a_2}{(16\pi)^3}\left(\frac{A}{4l_p^2}\right)^{-2}+ \frac{{a_2}^2 }{(16\pi)^4}\left(\frac{A}{4l_p^2}\right)^{-3}+\mathcal{O}({a_1}^2 a_2 , a_1 {a_2}^2 , {a_1}^3,{{a_2}^3 }).
\label{entropy in terms of area}
\end{eqnarray}
\noindent This completes our discussion of the effect of the GUP on the thermodynamic properties of the Schwarzschild black hole. In Figure 1, we present the plot of the entropy of the black hole vs the horizon area for the GUP case and compare it with the standard case.

\begin{figure}[H]
	\centering
\includegraphics[width=100 mm]{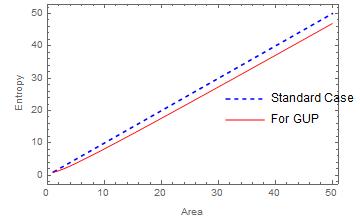}
\caption{The entropy vs area plot. Here the solid line (lower curve) represents the GUP case (considering only the first two terms in the right hand side of eq.(\ref{entropy in terms of area})), the dashed line (upper curve) represents the standard case.}
\end{figure}

\subsection{Energy output as a function of time}                                                                                                                                                                                                                                                                                                                                                                                                                                                                                                                                                                                                                                                                                                                                                                                                                                                                                                                                                                                                                                                                                                                                                               
Due to radiation of the black hole, the mass of the black hole reduces  while its temperature keeps on increasing. If one assumes that the energy loss is dominated by photons, then one can apply the Stefan-Boltzmann law to estimate the energy radiated as a function of time
\begin{eqnarray}
\frac{dM}{dt} = -\sigma A {T_H}^4
\label{Stefan-Boltzmann law 1}
\end{eqnarray}
where $\sigma$ is the Stefan-Boltzmann constant. In terms of Schwarzschild black hole mass M with the horizon area  $A=4\pi r_{s}^2 =\frac{16\pi G^2 M^2}{c^4}$, above equation implies
\begin{eqnarray}
\frac{d}{dt}\left(\frac{8\pi M}{M_p}\right) =
 -\sigma\frac{2 G^2 c^4 {M_p}^5}{{k_B}^4}\left(\frac{8\pi M}{M_p}\right)^2 {T'}^4
\label{Stefan-Boltzmann law Standerd 1}
\end{eqnarray}
where we have used $ T' = \frac{k_B T_H}{M_p c^2}$.

 \noindent We now write the above equation taking into account the effect of the GUP. Thus considering the mass-temperature relation ( \ref{Mass-Temp Relation3}), the radiation rate takes the following form

\begin{eqnarray}
\frac{dx}{dt} = -\frac{1}{t_{ch}x^2}\left[\frac{1}{x} + \frac{a_1}{{ x}^3} + \frac{a_2}{{x}^5} + \frac{2{a_1}^2}{{x}^5} + \frac{6 a_1 a_2}{{x}^7} + \frac{4{a_2}^2}{{x}^9}\right]^4
\label{Radiation Rate 2}
\end{eqnarray}
where we have set $x = \frac{8\pi M}{M_p}$ and the characteristic time $t_{ch}$ is being defined as  $t_{ch} = \frac{{k_B}^4}{2 \sigma {M_p}^5 c^4 G^2}$.
If $x_i$ refers to the initial mass at time $t=0$, the solution of 
the above equation yields the mass-time relation. Upto $ \mathcal{O}(a_1, a_2)$, we have
\begin{eqnarray}
x = \left[-\frac{3t}{t_{ch}}+ x_i^3 -12{a_1}{x_i}+ \frac{12 a_2}{x_i} +12{a_1}\left({x_i}^3 - \frac{3t}{t_{ch}}\right)^\frac{1}{3} -12{a_2}\left(x_i^3 - \frac{3t}{t_{ch}}\right)^
{-\frac{1}{3}} \right]^\frac{1}{3}
\end{eqnarray}
where 
\begin{eqnarray}
\frac{t}{t_{ch}} = \frac{{x_i}^3}{3}-4{a_1}{x_i}+\frac{4a_2}{x_i}.
\end{eqnarray}
\noindent In Figures 2 and 3 we have plotted the mass of the black hole as a function of time and the radiation rate  as a function of time.

\begin{figure}[H]
\centering
\begin{minipage}{.45\linewidth}
  \includegraphics[width=\linewidth]{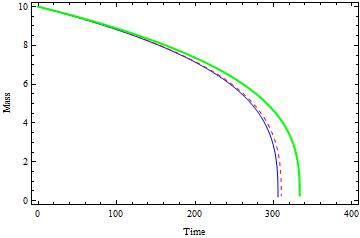}
  \caption{The mass of the black hole vs time. The mass is in units of Planck mass and the time is in units of characteristic time. Here thin line (lower curve) represents GUP case (considering both $a_2$ and $a_1$), the dashed line (middle curve) represents  GUP case (considering only $a_1$) and thick line (upper curve) represents the standard case.}
\end{minipage}
\hspace{.07\linewidth}
\begin{minipage}{.45\linewidth}
 \includegraphics[width=\linewidth]{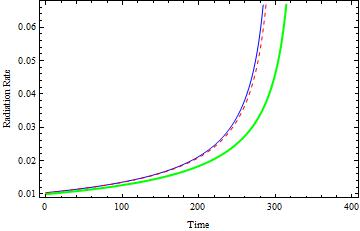}
  \caption{The Radiation Rate of the black hole vs time. The rate is in units of Planck mass per characteristic time and
  the time is in units of characteristic time.  Here thin line (upper curve) represents GUP case (considering both $a_2$ and $a_1$), the dashed line (middle curve) represents  GUP case (considering only $a_1$) and thick line (lower curve) represents the standard case.}
\end{minipage}
\end{figure}

\section{Thermodynamics of Reissner-Nordstr\"{o}m black hole}
\noindent  In this section, we consider the Reissner-Nordstr\"{o}m (RN) black hole of mass $M$ and charge $Q$.
In this case, near the horizon of the black hole, the position uncertainty of a particle will be of the order of the 
RN radius  of the black hole
\begin{eqnarray}
\Delta x&=&\epsilon r_h\nonumber\\
r_h &=& \frac{Gr_0}{c^2}\nonumber\\
r_0 &=& M+\sqrt{M^2 -Q^2}
\label{RN}
\end{eqnarray}
where $r_h$ is the radius of the horizon of the RN black hole. Substituting the value of $\Delta p$ and $\Delta x$ from eq.(\ref{momentum uncertainty-temp. reln.}) and eq.(\ref{RN}),
the GUP (\ref{saturaed form of GUP}) can be rewritten as
\begin{eqnarray}
r_0 = \frac{\hbar c^3}{2 \epsilon G k_B T}\left[1+ a_1\left(\frac{l_p \Delta p}{\hbar}\right)^2 + a_2\left(\frac{l_p \Delta p}{\hbar}\right)^4\right].
\label{RN Mass-Temp}
\end{eqnarray}
Once again, in the absence of correction due to GUP, eq.(\ref{RN Mass-Temp}) reduces to
\begin{eqnarray}
r_0=\frac{M_{p}^2 c^2}{2\epsilon k_{B}T}~.
\label{RN without quantum correction}
\end{eqnarray}
Comparing the above relation with the semi-classical Hawking temperature   
$T=\frac{M_{p}^{2}c^2 (Mr_0 -Q^2)}{2\pi k_{B}r_{0}^{3}}$, yields the
value of $\epsilon$ to be
\begin{eqnarray}
\epsilon=\frac{\pi r_{0}^2}{(Mr_0 -Q^2)}~.
\label{epsilon for RN}
\end{eqnarray}  
This finally fixes  the form of the mass-charge-temperature relation (\ref{RN Mass-Temp}) to be
\begin{eqnarray}
\frac{r_{0}^2}{(r_0 -M)}=\frac{M_{p}}{2\pi}\left[\frac{M_p c^2}{k_B T}+a_1\frac{k_{B}T}{M_p c^2} + a_2\left(\frac{k_{B}T}{M_p c^2}\right)^3\right]
\label{RN Mass-Temp 1}
\end{eqnarray}
where the identity
\begin{eqnarray}
\frac{r_{0}}{(Mr_0 -Q^2)}=\frac{1}{(r_0 -M)}~
\label{Identity 1}
\end{eqnarray}
 has been used.

\noindent The heat capacity of the black hole can now be calculated using relation (\ref{heat capacity defn.}) and eq.(\ref{RN Mass-Temp 1}):
\begin{eqnarray}
C=\frac{k_B (r_0 -M)^3}{2\pi {r_0}^2(2r_0 -3M)}\left[-\left(\frac{M_p c^2}{k_B T}\right)^2 +a_1 + 3a_2 \left(\frac{k_B T}{M_p c^2}\right)^2\right]~.
\label{RN Heat Capacity}
\end{eqnarray}
To express the heat capacity in terms of the mass, once again we make use of the relation (\ref{notation for mass and temp}) to recast
eq.(\ref{RN Mass-Temp 1}) in the form
\begin{eqnarray}
a_2 {T'}^4 + a_1 {T'}^2 -\frac{g(r_0)}{M_p}T' +1=0
\label{RN Mass-Temp 2}
\end{eqnarray}
where
\begin{eqnarray}
g(r_0)=\frac{2\pi r_{0}^2}{(r_0 -M)}\nonumber~.
\end{eqnarray}

\noindent Now to find out the temperature where the radiation process stops, we set C = 0. Eq.(\ref{RN Heat Capacity}) therefore yields
\begin{eqnarray}
3a_2 {T'}^4 + a_1 {T'}^2 -1 = 0
\label{RN temp at which C = 0}
\end{eqnarray}

\noindent from which solution of $T'$ reads 
\begin{eqnarray}
T' = \frac{1}{\sqrt{6a_2}}\left\{-a_1 + \sqrt{{a_1}^2 + 12a_2}\right\}^\frac{1}{2}
\label{RN maxm. temp.}
\end{eqnarray}
\noindent where the positive sign before the square root has been taken to reproduce the result corresponding to the  limit $a_2 \rightarrow 0$ \cite{abhijit2}.

\noindent The remnant mass can now be computed by substituting eq.(\ref{RN maxm. temp.}) in eq.(\ref{RN Mass-Temp 2}). This would then give

\begin{eqnarray}
\frac{g(r_0)}{M_p} = \frac{\sqrt{6a_2}}{9a_2\left\{-a_1 + \sqrt{{a_1}^2 + 12a_2}\right\}^\frac{1}{2}}\left[-{a_1}^2 + a_1 \sqrt{{a_1}^2 + 12a_2} + 12a_2\right].
\label{RN Remnant Mass considering both $a_1$ and $a_2$} 
\end{eqnarray}
Thus we finally obtain the following cubic equation for the remnant mass 
\begin{eqnarray}
\left(\frac{2 M_p Z}{\pi}\right) M_{rem}^3 - \left(\frac{M_p Z}{2 \pi}\right)^2 M_{rem}^2 - \left(\frac{2 M_p Z}{\pi}\right) Q^2 {M_{rem}}  + Q^4 +\left(\frac{M_p Z}{2 \pi}\right)^2 Q^2=0
\label{RN cubic expression for remnant mass}
\end{eqnarray}
where
\begin{eqnarray}
Z = \frac{\sqrt{6a_2}}{9a_2\left\{-a_1 + \sqrt{{a_1}^2 + 12a_2}\right\}^\frac{1}{2}}\left[-{a_1}^2 + a_1 \sqrt{{a_1}^2 + 12a_2} + 12a_2\right].
\label{Expression for $Z$} 
\end{eqnarray}
Solution of eq.(\ref{RN cubic expression for remnant mass}) gives us the exact expression of remnant mass for the RN black hole. This yields
\begin{eqnarray}
M_{rem}=\frac{ M_p Z}{24\pi}\left[1+\frac{\left(\frac{M_p^2 Z^2}{4\pi^2} +48 Q^2\right)}
{\left(\frac{B}{2}\right)^{1/3}}+\frac{4\pi^2}{M_p^2 Z^2}\left(\frac{B}{2}\right)^{1/3}\right]
\label{RN remnant mass}
\end{eqnarray}
where 
\begin{eqnarray}
B&=&\frac{M_p^6 Z^6}{32\pi^6}-18\frac{M_p^4 Z^4}{\pi^4}Q^2-108\frac{M_p^2 Z^2}{\pi^2}Q^4\nonumber\\&&+\sqrt{-\frac{27}{16}\frac{M_p^{10} Z^{10}}{\pi^{10}}Q^2-\frac{350}{27}\frac{M_p^8 Z^8}{\pi^8}Q^4-4968\frac{M_p^6 Z^6}{\pi^6}Q^6+11664\frac{M_p^4 Z^4}{\pi^4}Q^8}~.
\label{RN remnant mass B expression}
\end{eqnarray}
\noindent The above expression for the remnant mass reduces to the remnant mass for the Schwarzschild black hole (\ref{Remnant Mass})
in the $Q\rightarrow0$ limit.

\noindent Finally, we proceed to compute the entropy of the RN black hole. To do that, we first obtain an expression of $T' (T)$ from  eq.(\ref{RN Mass-Temp 2}) in terms of the mass and the charge of the RN black hole. This gives upto  $ \mathcal{O}({a_1}^2,{a_1 a_2},{a_2}^2)$
\begin{eqnarray}
T' = \frac{M_p}{g(r_0)}\left[1+a_1\left(\frac{M_p}{g(r_0)}\right)^2+a_2\left(\frac{M_p}{g(r_0)}\right)^4+2{a_1}^2\left(\frac{M_p}{g(r_0)}\right)^4+6a_1 a_2\left(\frac{M_p}{g(r_0)}\right)^6+4{a_2}^2 \left(\frac{M_p}{g(r_0)}\right)^8\right].
\label{RN Temp}
\end{eqnarray}

\noindent From this one now can calculate the entropy for the RN black hole using eq.(\ref{RN Heat Capacity}) and eq.(\ref{RN Temp}):   
\begin{eqnarray}
\frac{S}{k_B} &=& \frac{\pi r_{h}^2}{l_{p}^2} -\frac{a_1}{16 \pi}\ln\left(\frac{\pi r_{h}^2}{l_{p}^2}\right)-\left(\frac{a_1 Q^2}{2^3 M_p^2}-\frac{a_2}{2^8 \pi^2}\right)\left(\frac{\pi r_{h}^2}{l_{p}^2}\right)^{-1} - \left(\frac{a_2}{2^7 \pi M_p^2}-\frac{a_1 \pi Q^4}{2^5 M_p^5}\right)\left(\frac{\pi r_{h}^2}{l_{p}^2}\right)^{-2}\nonumber\\
&& - \left(\frac{a_2 Q^4}{2^7  M_p^4}\right)\left(\frac{\pi r_{h}^2}{l_{p}^2}\right)^{-3} - \left(\frac{a_2\pi Q^6}{2^8  M_p^6}\right)\left(\frac{\pi r_{h}^2}{l_{p}^2}\right)^{-4}
 -\frac{1}{5} \left(\frac{a_2\pi ^2 Q^8}{2^8  M_p^8}\right)\left(\frac{\pi r_{h}^2}{l_{p}^2}\right)^{-5}+\mathcal{O}({a_1}^2,{a_1 a_2},{a_2}^2)\nonumber\\
 &=& \frac{S_{BH}}{k_B} -\frac{a_1}{16 \pi}\ln\left(\frac{S_{BH}}{k_B}\right)-\left(\frac{a_1 Q^2}{2^3 M_p^2}-\frac{a_2}{2^8 \pi^2}\right)\left(\frac{S_{BH}}{k_B}\right)^{-1} - \left(\frac{a_2}{2^7 \pi M_p^2}-\frac{a_1 \pi Q^4}{2^5 M_p^5}\right)\left(\frac{S_{BH}}{k_B}\right)^{-2}\nonumber\\
 && - \left(\frac{a_2 Q^4}{2^7  M_p^4}\right)\left(\frac{S_{BH}}{k_B}\right)^{-3} - \left(\frac{a_2\pi Q^6}{2^8  M_p^6}\right)\left(\frac{S_{BH}}{k_B}\right)^{-4}
  -\frac{1}{5} \left(\frac{a_2\pi ^2 Q^8}{2^8  M_p^8}\right)\left(\frac{S_{BH}}{k_B}\right)^{-5}+\mathcal{O}({a_1}^2,{a_1 a_2},{a_2}^2)\nonumber\\
\end{eqnarray}
where  $\frac{S_{BH}}{k_B} =\frac{\pi r_{h}^2}{l_{p}^2}$ is the semi-classical 
Bekenstein-Hawking entropy for the RN black hole.
\noindent  In terms of the area of the horizon $A =4\pi r_{h}^2 =4l_{p}^2 \frac{S_{BH}}{k_B}$, the above equation can be written as
\begin{eqnarray}
\frac{S}{k_B} &=& \frac{A}{4l_p^2} -\frac{a_1}{16 \pi}\ln\left(\frac{A}{4l_p^2}\right)-\left(\frac{a_1 Q^2}{2^3 M_p^2}-\frac{a_2}{2^8 \pi^2}\right)\left(\frac{A}{4l_p^2}\right)^{-1} - \left(\frac{a_2}{2^7 \pi M_p^2}-\frac{a_1 \pi Q^4}{2^5 M_p^5}\right)\left(\frac{A}{4l_p^2}\right)^{-2}\nonumber\\
&& - \left(\frac{a_2 Q^4}{2^7  M_p^4}\right)\left(\frac{A}{4l_p^2}\right)^{-3} - \left(\frac{a_2\pi Q^6}{2^8  M_p^6}\right)\left(\frac{A}{4l_p^2}\right)^{-4}
 -\frac{1}{5} \left(\frac{a_2\pi ^2 Q^8}{2^8  M_p^8}\right)\left(\frac{A}{4l_p^2}\right)^{-5}+\mathcal{O}({a_1}^2,{a_1 a_2},{a_2}^2)\nonumber\\
\label{RN Area Rheorem}
\end{eqnarray}
which is the area theorem for the RN black hole with corrections from the GUP containing higher order terms in the momentum uncertainty. 

\noindent We would like to conclude this section by mentioning that in \cite{nozari7}, it has been pointed out that there is a part of the information (leaking out of the black hole due to Hawking radiation) related to non-thermal GUP correlations. This insight may be important to provide a solution for the well-known information loss paradox and is worth investigating in future. 

\section{Conclusions}
\noindent In this paper, we have investigated the modifications of the various thermodynamic properties of Schwarzschild and Reissner-Nordstr\"{o}m black holes using higher order momentum uncertainty terms in the GUP. We obtain the GUP modified mass-temperature relation. This then leads to the existence of a remnant mass thereby preventing the complete evaporation of the black hole. The expression for the remnant and critical masses have been obtained analytically. In this regard, we observe that analytical expressions for these masses can be obtained even if we keep terms of the order of $(\Delta p)^8$ in the momentum uncertainty. Beyond this it is no longer possible to obtain analytical expression for the critical and remnant masses. We also compute the mass and energy outputs as functions of time using the Stefan-Boltzman law. We observe that these expressions get modified from the standard case as well as the case where the simplest form of the GUP is used. The expression for the entropy exhibits the well known area theorem in terms of the horizon area in both cases upto leading order corrections from the GUP.

\section*{Acknowledgements}
S.G. acknowledges the support by DST SERB,  India under Start Up Research Grant (Young Scientist), File No.YSS/2014/000180. The authors would also like to thank the referee for very useful comments.

\noindent There is no conflict of interests with any funding agency regarding the publication of this manuscript.


\begin{thebibliography}{99}
\bibitem{maggiore} M. Maggiore, ``\textit{A Generalized Uncertainty Principle in Quantum Gravity}", Phys. Lett. B 304 (1993) 65-69.
\bibitem{gross1} D. J. Gross and P. F. Mende, ``\textit{The High-Energy Behavior of String Scattering Amplitudes}", Phys. Lett. B 197 (1987) 129-134. 
\bibitem{gross2} D. J. Gross and P. F. Mende, ``\textit{String Theory Beyond the Planck Scale}",  Nucl. Phys. B 303 (1988) 407-454.
\bibitem{str}D.~Amati, M.~Ciafaloni and G.~Veneziano, ``\textit{Can Space-Time Be Probed Below the String Size?}", Phys. Lett. B 216 (1989) 41-47 .
\bibitem{konishi} K. Konishi, G. Paffuti and P. Provero, ``\textit{Minimum Physical Length and the Generalized Uncertainty Principle in String Theory}", Phys. Lett. B 234 (1990) 276-284.
\bibitem{scardigli} F. Scardigli, ``\textit{Generalized Uncertainty Principle in Quantum Gravity from Micro-Black Hole Gedanken Experiment}", Phys. Lett. B 452 (1999) 39-44.
\bibitem{koni}K. Konishi, G. Paffuti and P. Provero, ``\textit{Minimum Physical Length and the Generalized Uncertainty Principle in String Theory}",  Phys. Lett. B 234 (1990) 276-284. 
\bibitem{magg2}M. Maggiore, ``\textit{The Algebraic structure of the generalized uncertainty principle}",  Phys. Lett. B 319 (1993) 83-86.
\bibitem{magg3} M. Maggiore, ``\textit{Quantum groups, gravity and the generalized uncertainty principle}", Phys. Rev. D 49 (1994) 5182-5187.
\bibitem{amelino2} G. Amelino-Camelia, M. Arzano and A. Procaccini, ``\textit{
Severe constraints on loop-quantum-gravity energy-momentum dispersion relation from black-hole area-entropy law}",  Phys. Rev. D 70 (2004) 107501.
\bibitem{amelino3} G. Amelino-Camelia, M. Arzano and A. Procaccini, ``\textit{A Glimpse at the flat-spacetime limit of quantum gravity using the Bekenstein argument in reverse}", Int. J. Mod. Phys. D 13 (2004) 2337-2343.
\bibitem{amelino4} Amelino-Camelia, M. Arzano, Y. Ling and G. Mandanica, ``\textit{
Black-hole thermodynamics with modified dispersion relations and generalized uncertainty principles}",  Class. Quant. Grav. 23 (2006) 2585-2606.
\bibitem{sadatian} S. D. Sadatian and H. Dareyni, ``\textit{Correction of Cardy–Verlinde formula for Fermions and Bosons with modified dispersion relation}", Annals Phys. 380 (2017) 71-77.
\bibitem{magueijo} J . Magueijo and L. Smolin, ``\textit{String theories with deformed energy momentum relations and a possible nontachyonic bosonic string}",   Phys. Rev. D 71 (2005) 026010.
\bibitem{das1}S.~Das and E.C.~Vagenas, ``\textit{Universality of Quantum Gravity Corrections}",  Phys. Rev. Lett. 101 (2008) 221301.
\bibitem{das3}S.~Das and E.C.~Vagenas, ``\textit{Phenomenological Implications of the Generalized Uncertainty Principle}",  Can. J. Phys. 87 (2009) 233-240.
\bibitem{das2}S.~Das and E.C.~Vagenas, ``\textit{Reply to `Comment on `Universality of Quantum Gravity Corrections}", Phys. Rev. Lett. 104 (2010) 119002.
\bibitem{das4}A.F.~Ali, S.~Das and E.C.~Vagenas, ``\textit{
Discreteness of Space from the Generalized Uncertainty Principle}", Phys. Lett. B 678 (2009) 497-499.
\bibitem{das5}S.~Das, E.C.~Vagenas and A.F.~Ali, ``\textit{Discreteness of Space from GUP II: Relativistic Wave Equations}",  Phys. Lett. B 690 (2010) 407-412, Erratum: Phys. Lett. B 692 (2010) 342-342.
\bibitem{das6}A.F.~Ali, S.~Das and E.C.~Vagenas, ``\textit{A proposal for testing Quantum Gravity in the lab}",  Phys. Rev. D 84 (2011) 044013.
\bibitem{gambini}R. Gambini, and J. Pullin, ``\textit{Nonstandard optics from quantum space-time}",  Phys. Rev. D 59 (1999) 124021.
\bibitem{alfaro} J. Alfaro, H.A. Morales-Tecotl and L.F. Urrutia, ``\textit{Quantum gravity corrections to neutrino propagation}", Phys. Rev. Lett. 84 (2000) 2318-2321.
\bibitem{amelino5} G. Amelino-Camelia, L. Smolin and A. Starodubtsev, ``\textit{Quantum symmetry, the cosmological constant and Planck scale phenomenology}",  Class. Quant. Grav. 21 (2004) 3095-3110.
\bibitem{girelli}  F. Girelli, E. R. Livine and D. Oriti, ``\textit{Deformed special relativity as an effective flat limit of quantum gravity}",  Nucl. Phys. B 708 (2005) 411-433.
\bibitem{mann} S. Das and R. B. Mann, ``\textit{Planck scale effects on some low energy quantum phenomena}", Phys. Lett. B 704 (2011) 596–599.
\bibitem{Das7}S. Basilakos, S. Das and E. C. Vagenas, ``\textit{Quantum Gravity Corrections and Entropy at the Planck time}",  JCAP 1009 (2010) 027.
\bibitem{Das8}W. Chemissany, S. Das, A. F. Ali and E. C. Vagenas, ``\textit{Effect of the Generalized Uncertainty Principle on Post-Inflation Preheating}", JCAP 1112 (2011) 017. 
\bibitem{adler} R. J. Adler, P. Chen and D. I. Santiago, ``\textit{Some heuristic semi-classical derivations of the
Planck length, the Hawking effect and the Unruh effect}",  Gen. Rel. Grav. 33 (2001) 2101-2108. 
\bibitem{nozari1} K. Nozari and S. H. Mehdipour, ``\textit{Gravitational Uncertainty and Black Hole Remnants}", Mod. Phys. Lett. A20 (2005) 2937-2948.
\bibitem{nozari2} K. Nozari and S. H. Mehdipour, ``\textit{Quantum-Corrected Black Hole Thermodynamics in Extra Dimensions}", Int. J. Mod. Phys. A21 (2006) 4979-4992.
\bibitem{rb} R. Banerjee and S. Ghosh, ``\textit{Generalised Uncertainty Principle, Remnant Mass and Singularity Problem in Black Hole Thermodynamics}", Phys. Lett. B 688 (2010) 224-229.
\bibitem{nozari3} K. Nozari and S. Saghafi, ``\textit{Natural Cutoffs and Quantum Tunneling from Black Hole Horizon}", JHEP 11 (2012) 005.
\bibitem{abhijit1} S. Gangopadhyay, A. Dutta and A. Saha, ``\textit{
Generalized uncertainty principle and black hole thermodynamics}",  Gen. Rel. Grav. 46 (2014) 1661. 
\bibitem{abhijit2} A. Dutta and S. Gangopadhyay, ``\textit{
Remnant mass and entropy of black holes and modified uncertainty principle}",  Gen. Rel. Grav. 46 (2014) 1747. 
\bibitem{abhijit3} S. Gangopadhyay, A. Dutta and M. Faizal, ``\textit{Constraints on the Generalized Uncertainty Principle from Black Hole Thermodynamics}",  Euro. Phys. Lett. 112 (2015) no.2, 20006.
\bibitem{anacleto} M. A. Anacleto, F. A. Brito, G. C. Luna, E. Passos and J. Spinelly,  ``\textit{Quantum-corrected finite entropy of noncommutative acoustic black holes}", Annals Phys. 362 (2015) 436-448.
\bibitem{abhijit4} A. Dutta and S. Gangopadhyay, ``\textit{Thermodynamics of black holes and the symmetric generalized uncertainty principle}",  Int. J. Theor. Phys. 55 (2016) no.6, 2746-2754. 
\bibitem{zhou} S. Zhou and Ge-Rui Chen, ``\textit{Corrected black hole thermodynamics in Damour–Ruffini’s method with generalized uncertainty principle}", Int. J. Mod. Phys. D 26 (2016) no.07, 1750062. 
\bibitem{feng} Z. W. Feng, H. L. Li, X. T. Zu and S. Z. Yang, ``\textit{Quantum corrections to the thermodynamics of Schwarzschild–Tangherlini black hole and the generalized uncertainty principle}", Eur. Phys. J. C 76 (2016) no.4, 212. 
\bibitem{dvali1} G. Dvali, S. Folkerts and C. Germani, ``\textit{Physics of Trans-Planckian Gravity}",  Phys. Rev. D 84 (2011) 024039.
\bibitem{dvali2}G. Dvali and C. Gomez, ``\textit{Ultra-High Energy Probes of Classicalization}", J. of Cosmo. and Astroparticle Phys. 1207 (2012) 015.
\bibitem{isi} M. Isi, J. Mureika and P. Nicolini, ``\textit{Self-Completeness and the Generalized Uncertainty Principle}", JHEP 1311 (2013) 139.
\bibitem{nozari4} K. Nozari and S. H. Mehdipour, ``\textit{Hawking Radiation as Quantum Tunneling from Noncommutative Schwarzschild Black Hole}", Class. Quant. Grav. 25 (2 008) 175015.
\bibitem{nozari5} K. Nozari and S. H. Mehdipour, ``\textit{Quantum Gravity and Recovery of Information in Black Hole Evaporation}" Europhys. Lett. 84 (2008) 20008.
\bibitem{dehghani} M. Dehghani, ``\textit{Hawking tunneling radiation of the spherically symmetric black holes at the Planck scale}", Astrophys. Space Sci. 357 (2015) no.2, 169.
\bibitem{sg} S. Gangopadhyay, ``\textit{Minimal length effects in black hole thermodynamics from tunneling formalism}", Int. J. Theor. Phys. 55 (2016) no.1, 617-624.
\bibitem{yang} S. Z. Yang, Z. W. Feng and H. L. Li, ``\textit{The Tunneling Radiation from Non-Stationary Spherical Symmetry Black Holes and the Hamilton-Jacobi Equation}", Int. J. Theor. Phys. 56 (2017) no.2, 546-553.
\bibitem{alasfer} L. Alasfar, S. M. Alsaleh, A. F. Ali and E. C. Vagenas, ``\textit{The GUP and quantum Raychaudhuri equation}",  arXiv:1706.06502.
\bibitem{li} G. P. Li, J. Pu, Q. Q. Jiang and X. T. Zu, ``\textit{An Application of Lorentz Invariance Violation in Black Hole Thermodynamics}",  Eur. Phys. J. C 77 (2017) no.10, 666.
\bibitem{nozari6}  K. Nozari and A. S. Sefiedgar ``\textit{Comparison of Approaches to Quantum Correction of Black Hole Thermodynamics}", Phys. Lett. B635 (2006) 156-160.
\bibitem{hawk1}S.W.~Hawking, ``\textit{Black hole explosions}", Nature 248 (1974) 30-31.
\bibitem{hawk2}S.W.~Hawking, ``\textit{Particle Creation by Black Holes}", Commun. Math. Phys. 43 (1975) 199-220, Erratum: Commun. Math. Phys. 46 (1976) 206.
\bibitem{nozari7} K. Nozari and S. H. Mehdipour, ``\textit{Parikh-Wilczek Tunneling from Noncommutative Higher Dimensional Black Holes}", JHEP 0903 (2009) 061.



\end{thebibliography}
\end{document}